\documentclass[aps,prl,reprint]{revtex4-2}
\usepackage{amsmath}
\usepackage{amssymb}
\usepackage{graphicx}
\usepackage{subfigure}
\usepackage[utf8]{inputenc}
\usepackage{amsthm}
\begin{document}
	\title{A phase transition in the Susceptible-Infected model on hypernetworks}
	\author{Gadi Fibich}
	\email{fibich@tau.ac.il}
	\author{Guy Rothmann}
	\email{guy86222@gmail.com}
	\affiliation{Tel Aviv University}
\begin{abstract}
	We derive the master equations for the Susceptible-Infected (SI) model on general hypernetworks with~$N$-body interactions. We solve these equations exactly for infinite~$d$-regular hypernetworks, and obtain an explicit solution for the expected infection level as a function of time. The solution shows that the epidemic spreads out to the entire population as~$t \to \infty$ if and only if the initial infection level exceeds a positive threshold value. This phase transition is a high-order interactions effect, which is absent with pairwise interactions. 
\end{abstract}
\maketitle
Spreading models on complex networks have been used to understand how diseases and information propagate through populations~\cite{keeling2005networks, pastor2001epidemic}. Traditional models represent all the interactions as pairwise contacts between individuals, neglecting the fact that many real-world interactions involve groups larger than two, such as meetings, social gatherings, or co-authorship networks~\cite{battiston2020networks}. In recent years there has been a growing interest in modeling high-order interactions using hypergraphs and simplicial complexes, which can capture multi-body interactions more accurately~\cite{petri2018simplicial, bick2021higher}.

The Susceptible-Infected-Susceptible (SIS) model and the Susceptible-Infected-Recovered (SIR) model on networks with pairwise interactions can exhibit non-trivial final states and phase transitions that are characterized by critical infection and recovery rates that distinguish disease-free from endemic states~\cite{pastor2015epidemic,moreno2002epidemic}. Moreover, in the SIS and SIR models on hypernetworks, high-order interactions can lead to critical phenomena that are absent in traditional pairwise interactions~\cite{iacopini2019simplicial, arruda2020social,PhysRevLett.132.087401}. 

In contrast, the Susceptible-Infected (SI) model on connected networks~\cite{fibich2024explicitsolutionssibass} and on hypernetworks~\cite{PhysRevE.110.054306} has so far shown a simpler spreading dynamics where for any initial infection level, the epidemic spreads out to the entire network. This can be attributed to the fact that the SI model only allows for a single unidirectional transition.
For example, in \cite{fibich2024explicitsolutionssibass}, the authors
obtained an exact explicit expression for the expected infection level as a function of time in the SI model on infinite~$d$-regular networks, which shows that for any positive initial infection level, the entire population becomes infected as~$t \to \infty$. 

In this Letter we extend the calculations of~\cite{fibich2024explicitsolutionssibass} to the SI model on hypernetworks with high-order interactions, and obtain an exact explicit solution for the expected infection level as a function of time. We show that the solution undergoes a phase transition, whereby the epidemic spreads to the entire population if and only if
the initial infection level exceeds a \emph{positive} threshold value. 
To our knowledge, this is the first demonstration that even in the "simple" SI model, high-order interactions can lead to critical phenomena which is absent with pairwise interactions.

\noindent\textit{The {\rm SI} model on~$N$-body Hypernetworks}\textemdash Consider~$M$~individuals~${\cal M}:=\{1, \dots, M\}$. The state of individual~$j$ at time~$t$ is
\begin{subequations}
	\label{Bass_SI_hypernetworks_general}
\begin{equation}
	X_j(t)=\begin{cases}
		1, \,\,{\rm if}\ j\ {\rm is \ infected \ at\ time}\ t,\\
		0, \,\, {\rm otherwise,}	
	\end{cases}
	\, j \in \cal M.
\end{equation}
The initial states at~$t=0$ are stochastic, so that
	\begin{equation}
		\label{eq:general_initial}
		X_j(0)=  X_j^0 \in \{0,1\}, \qquad j\in {\cal M},
	\end{equation}
	and
	\begin{equation}
		\label{1c}
		\mathbb{P}(X_j^0=1) =I^0, \quad
		\mathbb{P}(X_j^0=0) =1-I^0, \quad j\in\mathcal{M},
	\end{equation}
	where~$0<I^0<1$, and the random variables~$\{X_j^0 \}_{j \in \cal M}$ are independent. If~$j$ is susceptible, its infection time is piecewise exponentially distributed with the infection rate
	\begin{equation}
		\label{eq:lambda_j_3body(t)}
		\lambda_j(t) = \sum\limits_{\boldsymbol{k}\subset\mathcal{M}} q_{\boldsymbol{k}\rightarrow j} \prod_{i=1}^{N-1}X_{k_i}(t),\quad {\boldsymbol{k}:=\{k_i\}_{i=1}^{N-1}},
		\quad j \in {\cal M}.
	\end{equation}
	Here,~$q_{\boldsymbol{k}\rightarrow j} \geq 0$ is the infection rate of~$j$ due to the set of~$N-1$~nodes~${\boldsymbol{k}}$, provided that all the nodes in~$\boldsymbol{k}$ are infected. In addition,~$q_{\boldsymbol{k}\rightarrow j}>0$ if and only if~${j\not\in\boldsymbol{k}}$ and the directional hyperdge~$\boldsymbol{k}\rightarrow j$ exists. Once~$j$ becomes infected, it remains so at all later times.
\end{subequations}

The quantity of most interest is the expected infection level
\begin{equation}
	\label{eq:number_to_fraction-general}
	[I](t):=	\frac{1}{M} \sum_{j=1}^{M} [I_j](t), \qquad [I_j](t) :=\mathbb{E}[X_j](t),
\end{equation}
where~$[I_j]$ is the infection probability of node~$j$. To compute~$[I](t)$, let~$S_{\Omega}(t)$ denote the event the all the nodes in~$\Omega\subset\mathcal{M}$ are susceptible at time~$t$, and let~$[S_{\Omega}](t):=\mathbb{P}\bigl(S_{\Omega}(t)\bigr)$.
The stochastic dynamics of~\eqref{Bass_SI_hypernetworks_general} can be modeled by the master equations
\begin{subequations}
	\label{eq:master-equations}
\begin{equation}
\frac{d[S_{\Omega}]}{dt}=-\sum_{\boldsymbol{k}\subset\Omega^{c}, \, |\boldsymbol{k}|=N-1}q_{\boldsymbol{k}\rightarrow\Omega}[S_{\Omega}\cap I_{\boldsymbol{k}}],
\label{ME:eq1}
\end{equation}
 where~$\Omega^c:=\mathcal{M}\setminus\Omega$,~${q_{\boldsymbol{k}\rightarrow\Omega}:=\sum_{m\in\Omega}q_{\boldsymbol{k}\rightarrow m}}$ is the infection rate of the nodes in~$\Omega$ due to the~$N-1$ in~$\boldsymbol{k}$, and~$[S_{\Omega}\cap I_{\boldsymbol{k}}]$ is the probability that 
all the nodes in~${\Omega}$ are susceptible and 
all the nodes in~${\boldsymbol{k}}$ are infected.
 Using the inclusion–exclusion principle,~$[S_{\Omega}\cap I_{\boldsymbol{k}}]$ can be expressed as
\begin{equation}
	[S_{\Omega}\cap I_{\boldsymbol{k}}]=\sum_{i=0}^{N-1}(-1)^{i}\sum_{\boldsymbol{n}\subseteq\boldsymbol{k},\, |\boldsymbol{n}|=i}\Big[S_{\Omega\cup\boldsymbol{n}}\Big].
	\label{ME:eq2}
\end{equation} 
\end{subequations}
Combining~\eqref{ME:eq1} and~\eqref{ME:eq2}, we obtain
\begin{subequations}
	\label{ME_final}
\begin{equation}
\frac{d[S_{\Omega}]}{dt}=-\hspace{-0.6cm}\sum_{\boldsymbol{k}\subset\Omega^{c}, \, |\boldsymbol{k}|=N-1}q_{\boldsymbol{k}\rightarrow\Omega}\sum_{i=0}^{N-1}(-1)^{i}\sum_{\boldsymbol{n}\subseteq\boldsymbol{k},\, |\boldsymbol{n}|=i}\Big[S_{\Omega\cup\boldsymbol{n}}\Big].
	\label{ME:eq3}
\end{equation} 
The initial conditions are, see~\eqref{1c},
\begin{equation}
	\label{ME:eq4}
	[S_{\Omega}](0)=(1-I^0)^{\vert\Omega\vert}.
\end{equation}
\end{subequations}
The master equations~\eqref{ME_final} are
a closed system of~$2^M-1$ equations for~$\{[S_{\Omega}]\}_{\emptyset\neq\Omega\subset\mathcal{M}}$. 
\emph{These equations are exact, as they are derived without making any approximation}. For~$N=3$, the system~\eqref{ME_final} reduces to the one derived in~\cite{PhysRevE.110.054306}.
\\
\noindent\textit{$d$-regular~$N$-body hypernetworks}\textemdash An undirected~$N$-body hypergraph~$H$ is called \emph{$d$-regular} if every node has a hyperdegree~$d$.  Let~${\boldsymbol{E}=(e_{\boldsymbol{k},j})}$ be the adjacency tensor of~$H$, 
 and let all the hyperedges have weight~$\dfrac{q}{d}$. The corresponding~$d$-regular~$N$-body hypernetwork is given by
\begin{equation}
	\label{infinite_d_reg}
	q_{\boldsymbol{k}\rightarrow j}=\frac{q}{d}e_{\boldsymbol{k},j},\quad j\in\mathcal{M},\quad \boldsymbol{k}\subset\mathcal{M},\quad \vert\boldsymbol{k}\vert=N-1.
\end{equation}
As~$M\rightarrow\infty$, the master equations (\ref{ME_final},\ref{infinite_d_reg}) for infinite~$d$-regular~$N$-body hypernetwork have the exact explicit solution, see Supplementary Material~(SM),
\begin{subequations}
			\label{u_d_reg_ODE_hypergraph}
\begin{equation}
	\label{6a}
	[I^{\mathrm{d-reg}}_{\mathrm{N-body}}](t)=1-(1-I^{0})u^{d}\Big(\frac{qt}{d}\Big),
\end{equation}
where~$\frac{u(\cdot)}{1-I^0}$ is the susceptible probability of a degree-one node in an otherwise infinite~$d$-regular~$N$-body hypernetwork,
	\begin{equation}
		\label{6b}
\frac{du}{d\tau}=F^{\mathrm{d-reg}}_{\mathrm{N-body}}(u,I^0),\qquad u(0)=1,
	\end{equation}
and
	\begin{equation}
		\label{F_d_reg}
	F^{\mathrm{d-reg}}_{\mathrm{N-body}}:=1-u-\Big(1-(1-I^{0})u^{d-1}\Big)^{N-1}.
\end{equation}
\end{subequations}
On two-body networks ($N=2$),  the explicit solution~\eqref{u_d_reg_ODE_hypergraph} reduces to the one obtained in~\cite{fibich2024explicitsolutionssibass}. 

\begin{figure}[!ht]
	\centering
	\includegraphics[width=0.48\textwidth]{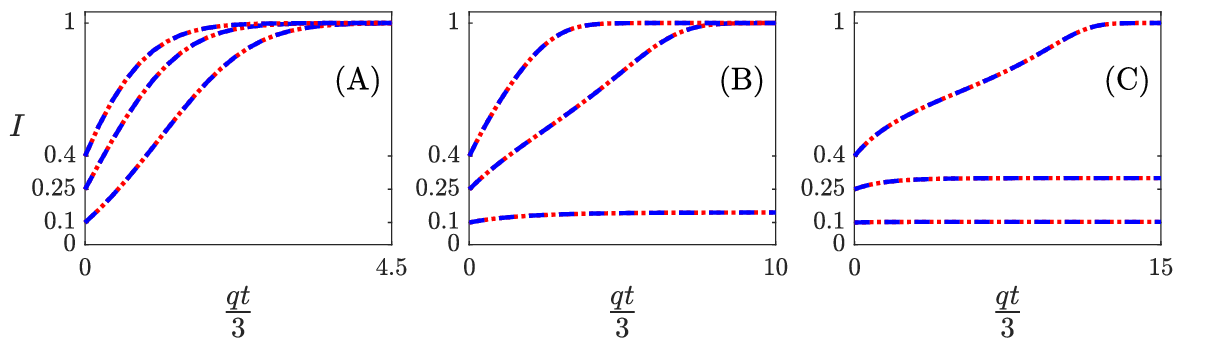} 
	\caption{The infection level on~$3$-regular~$N$-body hypernetworks as a function of time,
		for~$I^0 = 0.1, 0.25$, and~$0.4$.  The numerical solution of the SI model~(\ref{Bass_SI_hypernetworks_general},\ref{infinite_d_reg}) [red dots] is indistinguishable from the explicit solution~\eqref{u_d_reg_ODE_hypergraph} [blue dashes]. (A)~$N=2$. (B)~$N=3$. (C)~$N=4$.}
	\label{fig:f_d_reg_SI_vs_time}
\end{figure}

\begin{figure}[!ht]
	\centering
	\includegraphics[width=0.48\textwidth]{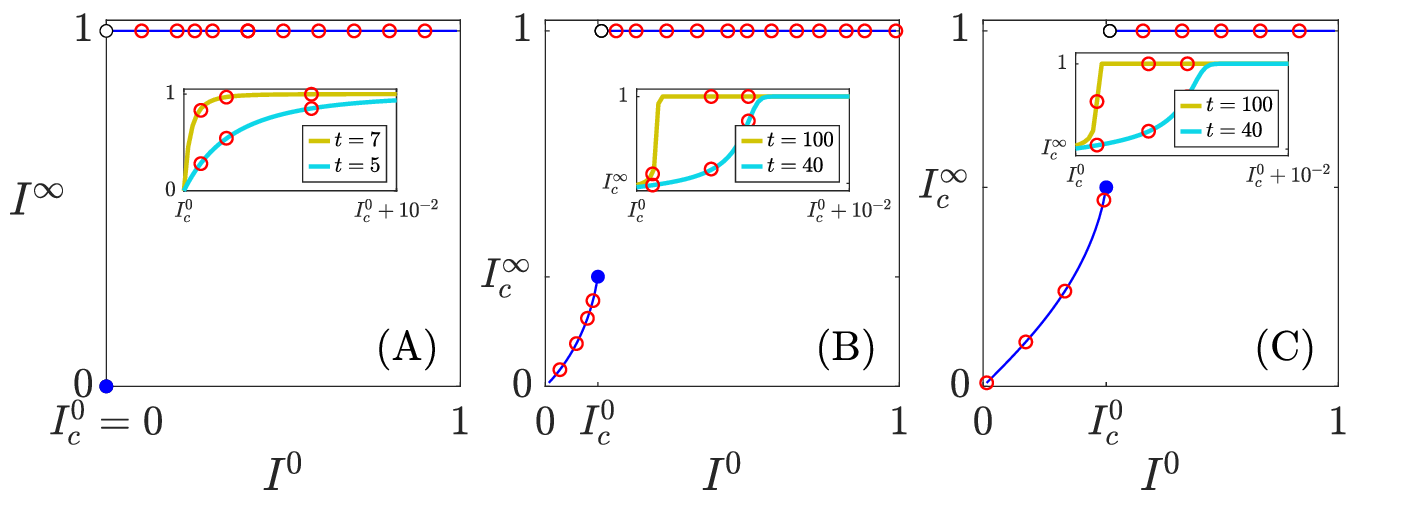} 
	\caption{The final infection level as a function of the initial infection level in the SI model on~$3$-regular~$N$-body hypernetworks. The red circles are numerical simulations of the SI model (\ref{Bass_SI_hypernetworks_general},\ref{infinite_d_reg}); The blue line is the explicit solution~\eqref{7}. The insets show the infection levels at~$t=t_1$ and at~$t=t_2$ for initial infection levels slightly above the critical threshold~$I^0_c$. 
	(A)~$N=2$,~$I^0_c=0$, $t_1=5$,~$t_2=7$. (B)~$N=3$,~$I^0_c=\frac{5}{32}$, $t_1=40$,~$t_2=100$. (C)~$N=4$,~$I^0_c=\frac{1}{108}(6+7\sqrt{21})$,~$t_1=40$,~$t_2=100$. }
	\label{fig:f_d_reg_SI_asymptotics}
\end{figure}

Fig.~\ref{fig:f_d_reg_SI_vs_time} confirms the excellent agreement between 
numerical simulations of the SI model~(\ref{Bass_SI_hypernetworks_general},\ref{infinite_d_reg}) on~$d$-regular networks and hypernetworks and the exact explicit solution~\eqref{u_d_reg_ODE_hypergraph}. 
When~$N=2$, the final infection level~$I^\infty:=\lim_{t\rightarrow \infty}[I^{\mathrm{d-reg}}_{\mathrm{N-body}}]$ is equal to one for~$I^0 = 0.1, 0.25,0.4$,
i.e., the infection spreads to the entire network. When~$N=3$ or~$4$, however,~$I^\infty=1$  only if~$I^0$ is sufficiently large. Indeed, plotting~$I^{\infty}$ as a function of~$I^{0}$ reveals a \emph{jump discontinuity} at~$I^0=I^0_c$, see Fig~\ref{fig:f_d_reg_SI_asymptotics}, where~$I^0_c=0$ on networks~($N=2$), and~$I^0_c>0$ on hypernetworks~($N\geq 3$).

To prove these numerical observations, we note that the critical points of~\eqref{6b} are obtained by equating~$F^{\mathrm{d-reg}}_{\mathrm{N-body}}$ to zero. 
Since~$u(0)=1$ and~$\frac{du}{d\tau}(0)=F^{\mathrm{d-reg}}_{\mathrm{N-body}}(1,I^0)<0$,~$u(\tau)$ is monotonically decreasing towards the first critical point below~$1$, which we shall denote by 
\begin{subequations}
	\label{7}
\begin{equation}
	\label{7a}
u^\infty(I^0):=\max\limits_{u <1} \{u \mid F^{\mathrm{d-reg}}_{\mathrm{N-body}}(u,I^0)=0 \}.
\end{equation}
 Thus,~$u^\infty:=\lim_{\tau\rightarrow\infty}u(\tau)$, and so by~\eqref{6a},
\begin{equation}
	\label{I_infty_d_reg}
	I^{\infty}=1-(1-I^0)(u^{\infty})^d.
\end{equation}
\end{subequations}  

\begin{figure}[!ht]
	\centering
	\includegraphics[width=0.48\textwidth]{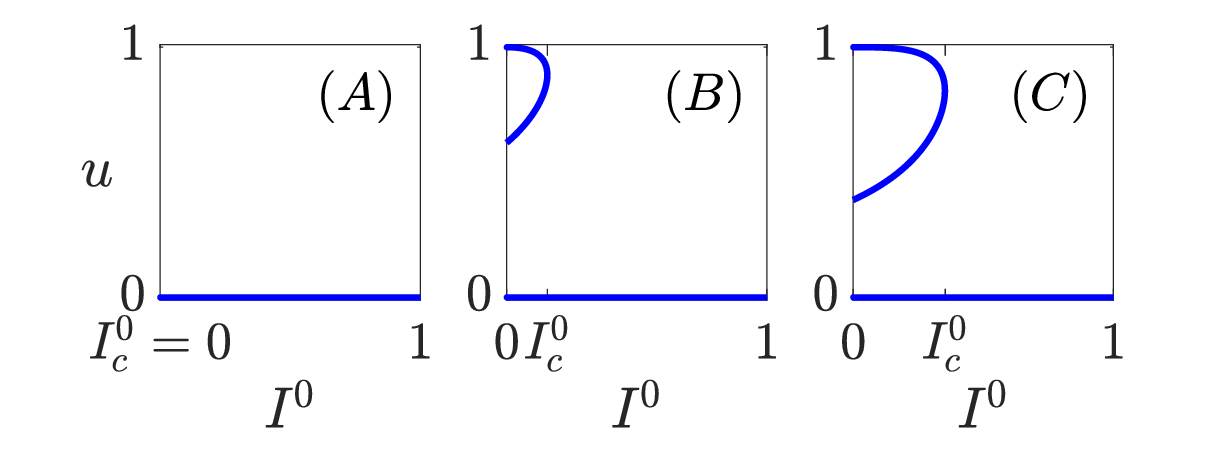} 
	\caption{ The critical points~$F^{3-\mathrm{reg}}_{\mathrm{N-body}}(u,I^0)=0$ (blue solid line) in~$[0,1]^2$. (A)~$N=2$,~$I^0_c=0$. (B)~$N=3$,~$I^0_c=\frac{5}{32}$. (C)~${N=4}$,~$I^0_c=\frac{1}{108}(6+7\sqrt{21})$.}
	\label{fig:Bifurcation_diagram}
\end{figure}

Specifically, on two-body networks~($N=2$),
\begin{equation*}
	F^{\mathrm{d-reg}}_{\mathrm{2-body}}:= u \Bigl((1-I^0)u^{d-2}-1\Big)<0, \qquad 0<u<1.
\end{equation*}
Therefore~$u^\infty\equiv0$ for any~$0<I^0<1$, see~Fig~\ref{fig:Bifurcation_diagram}A, and so
\begin{equation*}
	\label{I_infty_jump_N2}
	\begin{cases}
		I^{\infty}=0, & \mathrm{if} \,I^{0}=0,\\
		I^{\infty}=1, & \mathrm{if} \, 0<I^{0}\leq1.
	\end{cases}
\end{equation*}
Hence,~$I^0_c=0$,  see~Fig~\ref{fig:f_d_reg_SI_asymptotics}A.

On~$N$-body hypernetworks with~$N \ge 3$, however,
$F^{\mathrm{d-reg}}_{\mathrm{N-body}}=uP(u)$,
where~$P(u)$ is a polynomial of degree~$(N-1)(d-1)-1$ that has two real roots in~$(0,1)$ for~$0<I^0< I^0_c$, a double root at~$I^0_c$, and no real roots in~$(0,1)$ for~$I^0_c<I^0<1$, see SM and Fig~\ref{fig:Bifurcation_diagram}B and C. 
Therefore,
\begin{equation*}
	\label{I_infty_jump}
\begin{cases}
	I^{0}_c\leq I^{\infty}<1, & \mathrm{if} \, 0\leq I^{0}\leq I_{c}^{0},\\
	I^{\infty}=1, &  \mathrm{if} \,I_{c}^{0}<I^{0}\leq1.
\end{cases}
\end{equation*}
In particular,~$0<I^0_c<1$.
The dynamical system~\eqref{u_d_reg_ODE_hypergraph} thus admits a \emph{saddle-node bifurcation}, that leads to a phase transition at~$I_c^0$~(Fig~\ref{fig:Bifurcation_diagram}B and C). This results in a critical slowdown of the dynamics for~$I^0$ slightly above~$I^0_c$, see insets of Fig~\ref{fig:f_d_reg_SI_asymptotics}.
 
Intuitively, on infinite $d$-regular networks ($N=2$), any two nodes are connected by a finite path, with probability one~\cite{Frieze_Karoński_2015}. Hence, a single infected node at~$t=0$ is sufficient for the epidemic to spread out to the entire network as~$t\rightarrow\infty$, and so~$I^0_c=0$.
Similarly, on infinite $d$-regular $N$-body hypernetworks, any two nodes are connected by a finite hyperpath with probability one~\cite{Greenhill_Isaev_Liang_2022}. Maintaining the propagation of  an infection along a hyperpath, however, requires more than just for the first hyperedge to propagate the infection. Indeed, assume that $N-1$ nodes within a hyperedge are initially infected. Then the hyperedge propagates the infection to its~$N^{\rm th}$ node. To further propagate the infection to additional hyperedges,~$N-2$ nodes of the new hyperedge should be infected, in addition to the infected node from the previous hyperedge. Therefore, the initial infection level~$I^0$ should be sufficiently large, for the infection to be able to spread out to the entire hypernetwork.

The explicit expression~\eqref{7} for the final infection level~$I^{\infty}$ has the following interpretation.
As~$t\rightarrow\infty$, an arbitrary node~$j$ is susceptible (with probability~$1-I^{\infty}$) if and only if~$j$ was not infected initially~(with probability~$1-I^0$) and if none of its~$d$ hyperedges~$\{\boldsymbol{e_i}\}_{i=1}^d$ transmitted the infection to~$j$.
Let~$\theta$ denote the probability that a hyperedge~$\boldsymbol{e_i}$ that contains~$j$ has not transmitted the infection to~$j$ as~$t\rightarrow\infty$. Then 
\begin{subequations}
	\label{eq111ABC}
	\begin{equation}
		\label{eq111}
		1-I^{\infty}=(1-I^0)\theta^d.
	\end{equation}
Thus,~$\theta$ is the probability that a hyperedge~$\boldsymbol{e_i}$ that contains~$j$ has at most~$N-2$ infected nodes as~$t\rightarrow\infty$, conditioned on the event that~$j$ is susceptible as~$t\rightarrow0$. Let~$\phi$ denote that probability that a node~$k\in\boldsymbol{e_i}\setminus\{j\}$ is infected as~$t\rightarrow\infty$. Then
	\begin{equation}
		\label{eq112}
		\theta=1-\phi^{N-1}.
	\end{equation} 
The node~$k$ is susceptible as~$t\rightarrow\infty$ if and only if it was not infected initially, and if none of its other~$d-1$ hyperedges transmitted the infection to it. Therefore, 
	\begin{equation}
		\label{eq113}
		1-\phi = (1-I^0)\theta^{d-1}.
	\end{equation}
Substituting~\eqref{eq113} in~\eqref{eq112} and rearranging~\eqref{eq111} gives
\begin{equation*}
	\label{eq1030}
		I^{\infty}=1-(1-I^0)\theta^d,\qquad \theta=1-\bigl(1-(1-I^0)\theta^{d-1}\bigr)^{N-1},
\end{equation*}
which, after replacing~$\theta$ with~$u^{\infty}$, gives~\eqref{F_d_reg} and~\eqref{I_infty_d_reg}.
\end{subequations}

\begin{figure}[h]
	\centering
	\includegraphics[width=0.40\textwidth]{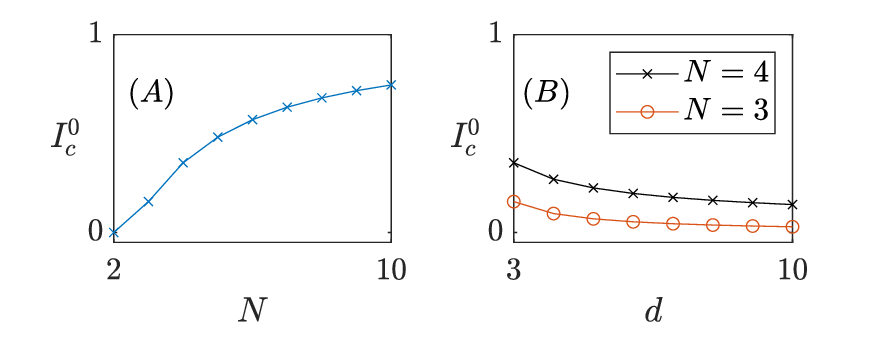} 
	\caption{(A)~The initial infection level threshold~$I^0_c$ as a function of~$N$ for~$d=3$. (B)~$I^0_c$ as a function of~$d$ for~$N=3$, see~\eqref{Ic0_explicit}, and~$N=4$, see~\eqref{Ic0_explicit_N_4}.}
	\label{fig:critical_threshold_of_d_N}
\end{figure}

The initial infection level threshold~$I_{c}^{0}$ increases as~$N$ increases and~$d$ is held fixed (Fig~\ref{fig:critical_threshold_of_d_N}A). This is because each hyperedge requires more infected nodes to propagate the infection.
Similarly,~$I_{c}^{0}$~decreases as~$d$ increases and~$N$ is held fixed, since the hypernetwork becomes more connected (Fig~\ref{fig:critical_threshold_of_d_N}B). Indeed, for~$N=3$, we can derive the explicit expression~(see SM)
 \begin{equation}
	\label{Ic0_explicit}
	I_{c}^{0}=1-(d-2)^{2-d}(d-1)^{1-d}(d-\frac{3}{2})^{2d-3}.
\end{equation}
The corresponding final infection level is, see~\eqref{6a}, 
\begin{equation*}
	\label{IC0inftyN3}
	I_{c}^{\infty}=\frac{4d^{2}-10d+5}{(2d-3)^{3}}.
\end{equation*}
Similarly, for~$N=4$ we have
\begin{subequations}
\label{Ic0_explicit_N_4}	
\begin{equation}
	\label{Ic0N4}
	I_{c}^{0}=1-(u_{\mathrm{max}})^{-(d-1)}(1-(1-u_{\mathrm{max}})^{\frac{1}{3}}),
\end{equation}
where
\begin{equation}
	u_{\mathrm{max}}=\frac{3(d-1)\bigl(18d(d-3)-\sqrt{12d-15}+39\bigr)}{2(3d-4)^{3}}.
\end{equation}
\end{subequations}

\noindent Furthermore, for any~$N\geq 3$ we have that, see SM,
\begin{equation}
	\label{IC0_asymptotics}
	I_{c}^{0}\sim\frac{N-2}{(N-1)^{\frac{N-1}{N-2}}}d^{-\frac{1}{N-2}},\qquad d\rightarrow\infty.
\end{equation}
As expected, the three explicit expressions~\eqref{Ic0_explicit},~\eqref{Ic0_explicit_N_4}, and~\eqref{IC0_asymptotics} are decreasing in~$d$.

In~\cite{BootstrapPercolation}, Morrison and Noel derived the asymptotic limit of the percolation threshold in infinite~$d$-regular~$N$-body hypergraphs that undergo a thinning process, whereby each hyperedge is independently kept with a probability that diminishes to zero as~$d\rightarrow\infty$. Remarkably, although their analysis does not cover the case of $d$-regular hypergraphs that do not undergo a thinning process, our asymptotic limit~\eqref{IC0_asymptotics} as~$d \to \infty$ precisely matches theirs.
Note, however, that in~\cite{BootstrapPercolation}, they did not obtain the exact critical infection level for finite~$d$, they did not derive the explicit expression~\eqref{u_d_reg_ODE_hypergraph} for the infection level as a function of time, and they did not show that~$I^{\infty}<1$ when~$I^0=I^0_c$.

In conclusion, previous work on the SI model on hypernetworks with~$N$-body interactions showed a dynamics which is qualitatively similar to that on networks~\cite{PhysRevE.110.054306}.  This Letter shows that high-order interactions can lead to a dramatic change in the dynamics, which is manifested by a phase transition at a \emph{positive} threshold, that is absent in networks with pairwise interactions. We expect that high-order interactions will lead to critical dynamics in other~$N$-body hypernetworks, such as sparse Erdős–Rényi hypernetworks. From a methodological perspective, this paper differs from most studies of the spreading dynamics on hypernetworks by deriving an~\emph{exact} expression for the infection level as a function of time, which is obtained by solving the master equations without making any approximation.

\noindent\textit{Acknowledgments}\textemdash 
We thank Michael Krivelevich, Wojciech Samotij, and Sahar Diskin for insightful discussions and valuable comments.

%\bibliographystyle{unsrt}  
%\bibliography{PRL.bib}
\end{document}

\section{Supplementary Material}
	\noindent\textit{Derivation of~\textup{(6)}}\textemdash 
	Following the same derivation as that of~\cite[Lemma 6.4]{fibich2024explicitsolutionssibass} with~$p=0$ gives
	\begin{subequations}
		\label{I_d_reg_N_body_derivation}
		\begin{equation}
			\label{eq4}
			[S^{\mathrm{d-reg}}_{\mathrm{N-body}}](t)=(1-I^{0})^{1-d}[\widetilde{S_j}]^d(t),
		\end{equation}
		where~$	[S^{\mathrm{d-reg}}_{\mathrm{N-body}}]$ is the expected susceptibility level in infinite~$d$-regular~$N$-body hypernetworks, and~$[\widetilde{S_j}]$ is the susceptibility probability of a node~$j$ in the modified hypernetwork~$\widetilde{\mathcal{N}}$, which is obtained by deleting~$d-1$ of the incoming hyperedges  to~$j$, and leaving only the  hyperedge~$\widetilde{\boldsymbol{k}}\rightarrow j$.
		By~(4),
		\begin{equation}
			\frac{d[\widetilde{S_{j}}]}{d\tau}+
			[\widetilde{S_{j}}]=-
			\sum_{i=1}^{N-1}(-1)^{i}\sum_{\boldsymbol{n}\subseteq \widetilde{\boldsymbol{k}}, \, |\boldsymbol{n}|=i}\Big[\widetilde{S_{\left\{ j\right\} \cup\boldsymbol{n}}}\Big],
			\label{eq:24}
		\end{equation}
		where~$[\widetilde{ \,\cdot \,}]$ denotes probabilities in the hypernetwork~$\widetilde{\mathcal{N}}$ and~${\tau:=q_{\widetilde{\boldsymbol{k}}\rightarrow j} t=\frac{q}{d}t}$. Let~$[\widetilde{\widetilde{\,\cdot\,}}]$ denote probabilities in the hypernetwork~$\widetilde{\widetilde{\mathcal{N}}}$, obtained from~$\widetilde{\mathcal{N}}$ by further deleting the hyperedge~$\widetilde{\boldsymbol{k}}\rightarrow j$ (thus making~$j$ an isolated node in~$\widetilde{\widetilde{\mathcal{N}}}$). By the indifference principle~\cite{Bass-boundary}, this deletion does not affect the susceptibility probability of the nodes in~$\widetilde{\boldsymbol{k}}$. Hence, 
	\end{subequations}
	\begin{subequations}
		\label{I_d_reg_N_body_derivation2}
		\begin{equation}
			[\widetilde{S_{\left\{ j\right\} \cup\boldsymbol{n}}}]
			=[\widetilde{\widetilde{S_{\left\{ j\right\} \cup\boldsymbol{n}}}}]
			=[\widetilde{\widetilde{S_{j}}}][\widetilde{\widetilde{S_{\boldsymbol{n}}}}]
			=[\widetilde{\widetilde{S_{j}}}][\widetilde{S_{\boldsymbol{n}}}].
			\label{eq:25}
		\end{equation}
		Since~$j$ is an isolated node in~$\widetilde{\widetilde{\mathcal{N}}}$,
		\begin{equation}
			[\widetilde{\widetilde{S_{j}}}]= 1-I^{0}.
		\end{equation}
		On infinite~$d$-regular~$N$-body hypernetworks, the infection events 
		by the nodes~$\boldsymbol{n}:=\{n_1,...,n_i\}$ are independent. Therefore,
		\begin{equation}
			[\widetilde{S_{\boldsymbol{n}}}]=\prod_{l=1}^{i}[\widetilde{S_{n_{l}}}].
			\label{eq:26}
		\end{equation}
		Each node~$n_l$ in the modified hypernetwork~$\widetilde{\mathcal{N}}$ has an hyperdegree~$d-1$, in an otherwise infinite~$d$-regular hypernetwork. Therefore, 
		by~\eqref{eq4}, with~$d$ replaced by~$(d-1)$,
		\begin{equation}
			[\widetilde{S_{n_l}}] = (1-I^{0})^{2-d}[\widetilde{S_j}]^{d-1}.
		\end{equation}
	\end{subequations}
	Combining \eqref{I_d_reg_N_body_derivation} and \eqref{I_d_reg_N_body_derivation2} and noting that there are~$\binom{N-1}{i}$ subsets~$\boldsymbol{n}$ of size~$i$ in a set of size~$N-1$, \eqref{eq:24} becomes
	\begin{equation*}
		\frac{d[\widetilde{S_{j}}]}{d\tau}+[\widetilde{S_{j}}]=
		(1-I^0)\sum_{i=1}^{N-1}\binom{N-1}{i}\Big(-(1-I^{0})^{2-d}[\widetilde{S_{j}}]^{d-1}\Big)^{i}.
	\end{equation*}
	Using the Binomial theorem and~$[\widetilde{S_{j}}]:=(1-I^0)u(t)$, we obtain~(6).
	
	\noindent\textit{The critical points of~\textup{(6)}}\textemdash 
	Let~$N,d\geq3$.
	The equation~$F^{\mathrm{d-reg}}_{\mathrm{N-body}}(u,I^0)=0$ can be solved for~$I^0$,  giving
	\begin{equation}
		I^0=f(u),\quad f(u):=1-u^{-a}(1-(1-u)^{b}),
		\label{I0f}
	\end{equation}
	where~$a=d-1\geq 2$ and~$b=\frac{1}{N-1}\leq\frac{1}{2}$.
	Now,~$f(0) = -\infty$ and~$f(1) = 0$. We claim that 
	there exists~$u_\mathrm{max}\in(0,1)$
	such that~$f(u)$ is increasing for~$0<u<  u_{\mathrm{max}}$
	and decreasing for~$ u_{\mathrm{max}}<u< 1$. Let 
	$I^0_c:=f(u_{\mathrm{max}})$. 
	Then~$0<I^0_c<1$, and  
	$F^{\mathrm{d-reg}}_{\mathrm{N-body}}(u;I^0)=0$ has two real roots in~$(0,1)$ for~$0<I^0< I^0_c$, a double root for~$I^0=I^0_c$, and no real roots in~$(0,1)$ for~$I^0_c<I^0<1$, as needed.  
	
	To prove the claim, we note that~$f'(u)=u^{-a-1}h(u)$, where
	$ h(u):=-a(1-u)^{b}+a-b(1-u)^{b-1}u$. Hence, 
	the sign of~$f'(u)$ and of~$h(u)$ are the same. Now,~$h(0)=0$ and 
	$h(1)=-\infty$. In addition, 
	$h'(u)=b(1-u)^{b-2}\big(bu+a(1-u)-1	\big)$
	vanishes in~$(0,1)$ only at~$u_0=\frac{a-1}{a-b}$. 
	Therefore,~$h(u)$ is monotonically increasing for~$0<u<u_0$ and 
	decreasing  for~$u_0<u<1$. Hence, there exists~$u_0<u_\mathrm{max}<1$,
	such that~$u_\mathrm{max}$ is the only root of~$h(u)=0$ in~$(0,1)$.
	Therefore,~$h(u)>0$  for~$0<u<u_\mathrm{max}$, and  
	$h(u)<0$  for~$u_\mathrm{max}<u<1$. Since the sign of~$f'(u)$ and of~$h(u)$ are the same, the claim follows. 
	\\
	\noindent\textit{The critical infection level~\textup{(9)} and~\textup{(10)}}\textemdash 
	Let~$N=3$ and~$d\geq3$. Since~$u_{\mathrm{max}}$ is the only local maximum of~$f$, the equation for~$u_{\mathrm{max}}$ is~$f'(u_\mathrm{max})=0$. It can be verified by direct substitution that~$f'(u)$ vanishes at~$u_\mathrm{max}:=\frac{(d-1)(d-2)}{(d-\frac{3}{2})^{2}}$, see~\eqref{I0f}. Substituting~$u_{\mathrm{max}}$ in~\eqref{I0f} yields~(9).
	For example, when~$N=d=3$,~$I^0_c=\frac{5}{32}$ and~$I^{\infty}_c=\frac{11}{27}$. 
	For~$N=4$ we used Mathematica to obtain the explicit solution of~$f'(u_{\mathrm{max}})=0$.
	\\
	\noindent\textit{Asymptotics of the critical infection level}\textemdash 
	Let~$N\geq3$ and~$d\rightarrow\infty$. The equation for~$u_{\mathrm{max}}$ is~$f'(u_{\mathrm{max}})=0$, and so by~\eqref{I0f},
	\begin{equation}
		\label{u_max_asymptotics}
		1-(1-u_{\mathrm{max}})^{b}=\frac{b}{a}u_{\mathrm{max}}(1-u_{\mathrm{max}})^{b-1}.
	\end{equation}
	Since~$0<b\leq\frac{1}{2}$ and~$0\leq u_{\mathrm{max}}\leq 1$, the left-hand side of~\eqref{u_max_asymptotics} remains bounded as~$d\rightarrow\infty$. Hence,~$u_{\mathrm{max}}\rightarrow1^{-}$  as~$a=d-1\rightarrow\infty$. Let~$x:=1-u_{\mathrm{max}}$. As~$x\rightarrow0^{+}$, equation~\eqref{u_max_asymptotics} becomes~$1\sim\frac{b}{a}x^{b-1}$, and so~$x\sim(\frac{a}{b})^{\frac{1}{b-1}}$. Using~\eqref{I0f} gives~$I_{c}^{0}\sim1-(1-x)^{-d}(1-x^{b})$. Substituting the expressions for~$a,b$ and~$x$ we get~$
	I_{c}^{0}\sim\bigl((N-1)^{-\frac{1}{N-2}}-(N-1)^{-\frac{N-1}{N-2}}\bigr)d^{-\frac{1}{N-2}}$. Using the identity~$(N-1)^{-\frac{1}{N-2}}-(N-1)^{-\frac{N-1}{N-2}}=\frac{N-2}{(N-1)^{\frac{N-1}{N-2}}}$, we obtain~(11).